\documentclass[twocolumn,showpacs,preprintnumbers,amsmath,amssymb]{revtex4}
\usepackage{graphics}
\usepackage{dcolumn}
\usepackage{bm}

\newcommand{\be}{\begin{equation}}
\newcommand{\ee}{\end{equation}}

\begin{document}

\title{Discrete-Time Quantum Field Theory and\\
the Deformed Super Virasoro Algebra}

\author{M. Chaichian}
\affiliation{High Energy Physics Division, Department of Physical
Sciences, University of Helsinki\\ and\\ Helsinki Institute of
Physics, P.O. Box 64, FIN-00014 Helsinki, Finland}

\author{P. Pre\v{s}najder}
\affiliation{Department of Theoretical Physics, Comenius
University\\ Mlynsk\'{a} dolina, SK-84248 Bratislava, Slovakia}

\today{}

\begin{abstract}

We show that the deformations of Virasoro and super Virasoro
algebra, constructed earlier on an abstract mathematical
background, emerge after Wick rotation, within an exact treatment
of discrete-time free field models on a circle. The deformation
parameter is $e^\lambda$, where $\lambda=\tau/\rho$ is the ratio
of the discrete-time scale $\tau$ and the radius $\rho$ of the
compact space.
\end{abstract}


\maketitle
\section{Introduction}

Over many years much attention has been paid to the Virasoro
algebra and super Virasoro algebra which play an important role in
conformal field theory and string theory (see, e.g., \cite{GSW},
\cite{Pol}). The super Virasoro algebra is an infinite Lie
superalgebra with even generators $B_n$ and $H_n$ (quadratic in
bosonic and fermionic oscillators respectively), and supplemented
by odd generators $G_r$ (linear in bosonic in bosonic and
fermionic oscillators). The index $n$ is integer, $r\in{\bf
Z}+1/2$ for Neveu-Schwarz sector, or $r\in{\bf Z}$ in Ramond
sector (Sugawara construction).

The deformations of the super Virasoro algebra are related to
deformations of this Sugawara construction: All even generators
$B^k_n$, $H^k_n$ and odd the ones $G^k_r$ are quadratic
expressions in deformed bosonic and fermionic oscillators. The
second index $k$ guarantees that $\{B^k_n\}$ closes to the double
indexed deformed Virasoro algebra \cite{CP1}. The supersymmetric
extension formed by $\{B^k_n, H^k_n, G^k_r\}$ was found in
\cite{BC} (the smallest set of $k$ for which these algebras close,
is $k=1,3,\dots$). When the deformation is removed the $k$
dependence becomes trivial.

Such deformations have been intensively studied in
\cite{Sat1}-\cite{KS}, mainly in connection with a formal
deformations in the conformal and/or string field theories. In
\cite{Sat2}-\cite{KS} it was shown that the second index $k$ is
closely related to the point-splitting of Virasoro currents and to
the deformation of the conformal symmetry.

Our construction in \cite{CP1} represents a particular realization
of the bosonization of Zamolodchikov-Faddeev algebras, which
proved to be a natural framework  for the deformed Virasoro
algebras \cite{LP} (for an overview see \cite{Od}). However, until
now meaning of the deformation, i.e. the physical interpretation
of the deformation parameter $\tau$ has not been clear. In this
paper we connect the parameter $\tau$ of the deformed super
Virasoro algebra with the discreteness of time in QFT.

In Sec. II, we describe first the free scalar field on a circle
subject to a standard continuous time evolution, in which
framework the Virasoro algebra appears naturally. Then we show how
the bosonic realization of the deformed Virasoro, proposed in
\cite{CP1}, emerges within discrete-time dynamics formulated in
\cite{Jar1}.

In Sec. III, we describe first the free fermionic field on a
circle within discrete-time approach modifying the approach
propsed in \cite{Jar1}. We recover the fermionic realization of
the deformed Virasoro algebra. Finally, the bosonic and fermionic
realizations are extended to the deformed super Virasoro algebra.

\section{Scalar field on a circle}

{\it Real time model.} The field action for a free massless real
scalar field on the circle $S^1$ with radius $\rho$ is defined by
\be S[\Phi ]\ =\ {1\over 4\pi\rho}\int_{{\bf R}\times S^1} dt
d\varphi [(\rho\partial_t\Phi )^2-(\partial_\varphi \Phi )^2 ]\
.\label{II.1}\ee
We can expand the field $\Phi (t,\varphi)=\Phi^*(t,\varphi)$ into
the Fourier modes
\be \Phi (t,\varphi)\ =\ \sum_{k>0}[c_k(t)e^{ik\varphi}+
e^{-ik\varphi}c^*_k (t)]\ ,\ k-{\rm integer}\ .\label{II.2}\ee
Inserting (\ref{II.2}) into the action, we obtain \be S[\Phi ]\ =\
\int_{\bf R}{dt\over\rho}\, \sum_{k>0}[\rho^2{\dot c}^*_k (t){\dot
c}_k(t) -k^2 c^*_k (t)c_k(t)]\ .\label{II.3}\ee
The canonically conjugate momentum to the modes $c_k(t)$ and
$c^*_k(t)$ are $\pi_k(t)=\rho {\dot c}^*_k(t)$ and
$\pi^*_k(t)=\rho{\dot c}_k (t)$, respectively. Solving the
corresponding equations of motion we obtain the solution
\be \Phi (t,\varphi )\ =\ \frac{i}{\sqrt{2}} \sum_{k\neq
0}\frac{1}{k} (a_ke^{-ik(t-\rho\varphi)/\rho}\ +\ b_k
e^{-ik(t+\rho\varphi)/\rho})\ .\label{II.4}\ee
Here we have used the notation $a_{-k}=a^*_k$ and $b_{-k}=b^*_k$.
The terms with expansion coefficients $a_k$ are interpreted as the
right-movers on a closed bosonic string, whereas those solutions
with $b_k$ as the left-movers. They are independent, and we can
treat them separately.

The equal-time canonical commutation relations
\be [c_k(t),\pi_{k'}(t)]\ =\ i\delta_{kk'}\ =\
[c^*_k(t),\pi^*_{k'} (t)]\, ,\ \ k,k'>0\ , \label{II.5}\ee
(with all other elementary commutators vanishing) are satisfied if
the coefficients $a_k$ and $b_k$ are replaced by two independent
infinite set of bosonic oscillators satisfying the commutation
relations:
\[ [a_k ,a_{k'}]\ =\ [b_k ,b_{k'}]\ =\
k\delta_{k+k',0}\ ,\]
\be [a_k,b_{k'}]\ =\ 0\ ,\ k,k'\neq 0\ .\label{II.5a}\ee

Next we collect essential steps of the discrete-time approach
proposed in \cite{Jar1}. In this approach the trajectory $q(t)$ is
replaced by a finite set of variables $q_n$, $n=0,1,\dots ,N$,
interpreted as positions at the given discrete-times
$t_n:q_n=q(t_i+ n\tau)$, $\tau =(t_f-t_i)/N$. The action integral
is replaced by the finite sum
\be S_\tau[q]\ =\ \sum_{n=0}^{N-1} S(q_n,q_{n+1})\
.\label{II.6}\ee
The function $S_n\equiv S(q_n,q_{n+1})$, called the system
function, specifies the dynamics of the system in question. The
equations of of motion have the form, \cite{Jar1}:
\be
\partial_{q_n}S_\tau[q]\ =\ \partial_{q_n}[S_{n-1}+S_n]\ = \ 0 \ ,\
\ n=1,\dots ,N-1\ .\label{II.7}\ee
A natural choice for $S_n$ can be the time-slice Hamilton
principal function $S_n=S_c (q_n,q_{n+1})$, i.e. the value of the
action integral calculated for the classical path $q(t)$ starting
at the given point $q_n=q(t_n)$ and terminating at
$q_{n+1}=q(t_{n+1})$.

For systems quadratic in positions and momenta, like harmonic
oscillator, the problem of quantization is reduced to the standard
one. Namely, the discrete-time dynamical variables $q_n$ and $p_n$
should be replaced by operators satisfying the canonical
"equal-time" commutations relations: $[q_n,p_n]=i$. For such
systems the momentum $p_n$ conjugated to $q_n$ can be defined as
the product of particle mass $\mu$ and the "discrete-time
velocity":
\be p_n\ =\ \frac{\mu}{ 2\tau}(q_{n+1}-q_{n-1})\ ,\label{7b}\ee
see e.g. \cite{Jar1}-\cite{Has}. The canonical commutation
relation then reads:
\be [q_n,{\mu\over 2\tau} (q_{n+1}-q_{n-1})]\ =\ i\
.\label{II.7a}\ee

Let us now consider the harmonic oscillator described by the
Lagrangian
\be L(q,{\dot q})\ =\ {1\over 2}\mu^2{\dot q}^2 -
{1\over 2}\mu^2\omega^2q^2\ .\label{II.8}\ee
The corresponding Hamilton principal function in variables
$q_n=q(n\tau )$ is:
\be S_c(q_n,q_{n+1} )\ =\ {\mu\omega\over
2\sin\omega\tau} [(q^2_{n+1} +q^2_n)\cos\omega \tau\, -\,
2q_{n+1}q_n]\ .\label{II.9}\ee
The solutions of equations of motion are
\be q_n\ =\ \frac{i}{\omega}\sqrt{\frac{\omega\tau}
{2\sin\omega\tau}} (e^{-in\omega\tau}a-e^{in\omega\tau}a^*)\
.\label{II.10}\ee
The quantization can be now performed directly. Replacing $a$ and
$a^*$ by annihilation and creation operators satisfying
\be [a,a^*]\ =\ {\sin\omega\tau\over\mu\tau}\ ,\label{II.11}\ee
the canonical commutation relation (\ref{II.7a}) follows directly.

Let us now apply this results to the free scalar field on a circle
with radius $\rho$. Putting $c_k(t)=(1/\sqrt{2})
[c^R_k(t)+ic^I_k(t)]$, the formula (\ref{II.3}) represents an
action for two infinite sets of harmonic oscillators $c^R_k={\rm
Re}c_k$ and $c^I_k={\rm Im}c_k$ with frequencies $\omega =k/\rho$,
$k$ - positive integer. The corresponding time-slice principal
Hamilton function is
$$ S_c(c^k_n,c^k_{n+1})\ =\ \sum_{k>0}{k\over 2\sin k\lambda}
[(c^k_nc^{k*}_n+c^k_{n+1}c^{k*}_{n+1})\cos k\lambda $$
\be -c^{k*}_nc^k_{n+1}-c^{k*}_{n+1}c^k_n ]\ .\label{II.12}\ee
Here $\lambda =\tau/\rho$ is the natural dimensionless parameter
induced by the discreteness of the time.

The quantum discrete-time version is obtained straight\-forwardly,
by repeating for any oscillator the steps which led from the
Lagrangian (\ref{II.8}) to the solutions (\ref{II.10}) given in
terms of annihilation and creation operators satisfying
(\ref{II.11}). Performing this procedure, we obtain the
discrete-time fields on a circle
\be \Phi_n(\varphi )\ =\ {i\over\sqrt{2}}\sum_{k\neq 0}
{\lambda\over\sin k\lambda} (a_ke^{-ik(n\lambda -\varphi)}\ +\ b_k
e^{-ik(n\lambda +\varphi)})\ ,\label{II.13}\ee
with annihilation and creation operators satisfying the deformed
commutation relations
\[ [a_k ,a_{k'}]\ =\ [b_k ,b_{k'}]\ =\ {\sin
k\lambda\over\lambda} \delta_{k+k',0}\ ,\]
\be [a_k ,b_{k'}]\ =\ 0\ ,\ k,k'\neq 0\ .\label{II.14}\ee
Here, $a_-k=a^*_k$, $b_{-k}=b^*_k$. The operators $a_k,k>0$, can
be interpreted as annihilation operator, provided $\sin
(k\lambda)>0$. This is guaranteed if the admissible values of $k$
are restricted to $0<k<{\pi\over\lambda}$.\\

{\it Euclidean time model.} In order to analyze the Euclidean
case, we have to substitute the time $t$ by $-it$; as a result the
the Euclidean action is defined by
\be S[\Phi ]\ =\ {1\over 4\pi\rho} \int_{{\bf R}\times S^1}
dtd\varphi [(\rho\partial_t\Phi )^2 +(\partial_\varphi\Phi )^2]\
,\label{III.1}\ee
with the field
\be \Phi (t,\varphi )\ =\ \sum_{k>0} [c_k(t)
e^{ik\varphi}+e^{-ik\varphi} c^*_k (-t)]\ ,\label{III.2}\ee
satisfying the Euclidean reality condition, $\Phi^\dagger
(t,\varphi )\equiv\Phi^* (-t,\varphi )=\Phi (t,\varphi )$. The
solution corresponding Euler-Lagrange equations in terms of the
variables $z=e^{(t+i\rho\varphi)/\rho}$ and ${\bar
z}=e^{(t-i\rho\varphi)/ \rho}$, reads
\be \Phi (z,{\bar z})\ =\ {i\over\sqrt 2} \sum_{k\neq 0}{1\over
k}(a_k z^{-k}\ +\ b_k {\bar z}^{-k} )\ \equiv\ \Phi
(z)+{\tilde\Phi}({\bar z})\ .\label{III.3}\ee
The canonical equal-time commutation relations are satisfied
provided that $a_k$ and $b_k$, $k\neq 0$, satisfy commutation
relations (\ref{II.5a}).

The Euclidean discrete-time version is obtained by analytic
continuation, \cite{Jar1}: the discrete-time step $\tau$ is
replaced by $-i\tau$. Performing this, the Euclidean Hamilton
principal function is:
$$ S_c(c^k_n,c^k_{n+1}) \ =\ \sum_{k>0}{k\over 2\sinh k\lambda}
[(c^k_nc^{k*}_n\ + \ c^k_{n+1}c^{k*}_{n+1})\cosh k\lambda $$
\be -\ c^k_{n+1}c^{k*}_n\ -\ c^{k*}_{n+1}c^{k*}_n ]\
.\label{III.4}\ee

The solution of equations of motion for the field $\Phi_n(\varphi
)\equiv\Phi (z,{\bar z})$ possesses in the variables
$z=e^{n\lambda +i\varphi}$ and ${\bar z}=e^{n\lambda -i\varphi}$
the mode expansion
\be \Phi (z,{\bar z})\ =\ {i\over\sqrt 2} \sum_{k\neq 0}{1\over
[k]_-}(a_k z^{-k}\ +\ b_k {\bar z}^{-k})\ \equiv\ \Phi
(z)+{\tilde\Phi}({\bar z})\ ,\label{III.6}\ee
where $[k]_-=(1/\lambda )\sinh k$. The Euclidean reality condition
$\Phi_n^\dagger (\varphi )\equiv\Phi_{-n}^* (\varphi )
=\Phi_n(\varphi )$ is satisfied provided $a_{-k}=a^*_k$ and
$b_{-k}=b^*_k$. The canonical commutation relations among fields
$\Phi (z,{\bar z})$ and field momenta
\be \Pi (z,{\bar z})\ =\ {1\over\sqrt 2}\sum_{k\neq 0}(a_k z^{-k}\
+\ b_k {\bar z}^{-k} )\ . \label{III.5}\ee
induce the following commutation relations for the oscillator
pairs:
\be [a_k ,b_{k'} ]\ =\ 0\ ,\ [a_k ,a_{k'}]\ =\ [b_k ,b_{k'}]\ =\
[k]_- \delta_{k+k',0}\ . \label{III.7}\ee
Since $[k]_-\neq 0$ for any $k>0$, there is no restriction on
allowed range of $k$. This is different from the real
discrete-time case.

The deformed Virasoro algebra generators can be expressed as
contour integrals (over circle in the complex with a given radius
$r=e^{l\lambda}$, $l$-positive integer):
\[ B^k_n \ =\ \oint{dz\over 2\pi i}\, z^n:\Pi (e^{k\lambda/2}z)\Pi
(e^{-k\lambda/2}z): \]
\be =\ {1\over 2} \sum_j[k({n\over 2}-j)]_+ :a_j a_{n-j}:\
,\label{III.8}\ee
where $\Pi (z)={1\over\sqrt 2}\sum a_k z^{-k}$ is the holomorphic
part of the field momentum $\Pi (z,{\bar z})$.
The anti-holomorphic part ${\tilde\Pi}({\bar z})={1\over\sqrt
2}\sum b_k {\bar z}^{-k}$ gives rise to another set of Virasoro
generators expressed in terms of $b_k$.

\section{Supersymmetric extension}

{\it Fermionic oscillator.} Let us consider the continuous time
fermionic Lagrangian
\be L\ =\ {i\over 2} \xi^*{\dot \xi} - {i\over 2}{\dot\xi}^*\xi
-\omega\xi^*\xi\ ,\label{IV.1}\ee depending on complex Grassmann
variables $\xi$ and $\xi^*$ and a parameter $\omega$ either
positive or negative. Since $L$ is linear in velocities, we can
specify only two of the four values $\xi_i, \xi_i^*$ and
$\xi_f,\xi_f^*$ at initial and final times.

The field action which for fixed $\xi_f^*$ and $\xi_i$ leads to
equations of motion, is
\be S[\xi^*,\xi ]\ =\ \int dt\, [{i\over 2} \xi^* {\dot\xi} -
{i\over 2}{\dot\xi}^*\xi -\omega\xi^*\xi]\ +\ {i\over
2}\xi_f^*\xi_f\ +\ {i\over 2}\xi_i^*\xi_i\ .\label{IV.2}\ee
Inserting here the solutions $\xi\ =\ e^{-i\omega (t-t_i)}\xi_i$,
$\xi^*\ =\ e^{-i\omega (t_f-t)}\xi_f^*$ of equations of motion, we
obtain the principal Hamilton function $S_c(\xi_f^*, \xi_i)=
ie^{-i\omega (t_f-t_i)}\xi_f^*\xi_i$. Alternatively, for fixed
$\xi_i^*$ and $\xi_f$ the proper field action
\be S[\xi^*, \xi]\ =\ \int dt\, [{i\over 2}\xi^*{\dot\xi} -
{i\over 2}{\dot\xi}^*\xi -\omega\xi^*\xi]\ -\ {i\over
2}\xi_f^*\xi_f\ -\ {i\over 2}\xi_i^*\xi_i\ .\label{IV.4}\ee
leads to the principal Hamilton function $S_c(\xi_i^*,\xi_f)=
ie^{i\omega (t_f-t_i)}\xi_f \xi_i^*$.

The Lagrangian (\ref{IV.1}) is singular and gives rise to first
class constraints $\eta (t)+{i\over 2}\xi^*(t)=0$ and $\eta^*(t)
-{i\over 2}\xi(t)=0$ among fermionic coordinates $\xi =\xi (t)$,
$\xi^*=\xi^*(t)$ and corresponding momenta $\eta =\eta (t)$,
$\eta^*= \eta^*(t)$. They lead to Dirac brackets
\[ \{\xi (t),\xi (t)\}_D = \{\xi^*(t),\xi^*(t)\}_D = 0\ ,\
\{\xi (t),\xi^*(t)\}_D = 1\ ,\]
which after quantization are replaced by anticommutation relations
\be [\xi (t),\xi (t)]\ =\ [\xi^*(t),\xi^*(t)]\ =\ 0\ ,\ [\xi
(t),\xi^*(t)]\ =\ 1\ .\label{IV.7}\ee

In the discrete-time case as dynamical variables we take $\xi_n
\equiv\xi (n\tau)$ for $n$-odd, and $\xi^*_{n'}\equiv
\xi^*(n'\tau)$ for $n'$-even (or equivalently, with the role of
$n$ even and odd interchanged). For the time-slice system function
we can take either $S_c(\xi_{n+1}^*,\xi_n)+S_c(\xi_{n-1}^*,\xi_n)$
or $S_c(\xi_{n'}^*,\xi_{n'-1})+S_c(\xi_{n'}^*,\xi_{n'+1})$. Both
choices give the same expression
\[ S_\tau\ =\ i\sum_n (\xi_{n+1}^*\xi_ne^{-i\omega\tau}-
\xi_{n-1}^* \xi_ne^{i\omega \tau})\]
\be =\ i\sum_{n'} (\xi_{n'}^*\xi_{n'-1}
e^{-i\omega\tau}-\xi_{n'}^* \xi_{n'+1} e^{i\omega \tau})\
.\label{IV.8}\ee
The corresponding equations of motion
\[ \xi_{n'+1}e^{i\omega \tau}\ =\ \xi_{n'-1}e^{-i\omega \tau}\ ,\
\xi^*_{n+1}e^{-i\omega \tau}\ =\ \xi^*_{n-1}e^{i\omega \tau}\ ,\]
have the solution
\be \xi_n\ =\ {1\over\cos\omega\tau} e^{-in\omega\tau}b\ ,\
\xi^*_{n'}\ =\ {1\over\cos\omega\tau} e^{in'\omega \tau}b^*\
.\label{IV.9}\ee

In the discrete-time case we replace (\ref{IV.7}) by
anticommutators
\[ [\xi_n ,\xi_n ]\ =\ [\xi^*_{n'} ,\xi^*_{n'}]\ =\ 0\ ,\]
\be [\xi_n ,{1\over 2}(\xi^*_{n+1}+\xi^*_{n-1})]\ =\ 1\ ,\ {\rm
or}\ \ [\xi^*_{n'} ,{1\over 2}(\xi_{n'+1}+\xi_{n'-1})]\ =\ 1\
.\label{IV.10}\ee
Here we have taken into account that $\xi_n$ is affiliated with
$n$-odd, whereas $\xi^*_{n'}$ with $n'$-even, i.e. $\xi^*_{n'}$ is
a link variable with respect to $\xi_n$, and vice versa. The
second line in (\ref{IV.10}) indicates both alternative choices
(in the first (second) anticommutator we replaced $\xi^*_{n'}$
($\xi_n$) by the nearest neighbor average). Inserting here
solutions (\ref{IV.9}) we find that the anticommutation relations
(\ref{IV.10}) are satisfied, for both choices, provided that $b$
and $b^*$ satisfy anticommutation relations
\be [b,b]\ =\ [b^*,b^*]\ =\ 0\ ,\ [b,b^*]\ =\ \cos\omega\tau\
.\label{IV.11}\ee
The left-hand-side of the last anticommutator represents a
positive operator. Therefore, we require $\cos\omega\tau >0$, i.e.
$-\pi /2<\omega\tau <+\pi /2$. For $\omega >0$ we interpret $b$ as
an annihilation operator and $b^*$ as a creation one, for $\omega
<0$ their interpretation is reversed.

The Euclidean version is obtained by the replacement $\tau\to
-i\tau$. Repeating all the steps leading to (\ref{IV.11}), we
obtain a fermionic oscillator pair satisfying the anticommutation
relations
\be [b,b]\ =\ [b^*,b^*]\ =\ 0\ ,\ [b,b^*]\ =\ \cosh\omega\tau\
.\label{IV.12}\ee
In this case there is no restriction on admissible values of
$\omega\tau $.\\

{\it Fermionic field on a circle}. The real time free fermio\-nic
field on a circle in Neveu-Schwarz sector satisfies antiperiodic
boundary conditions $\Psi (t,\varphi +2\pi) =-\Psi (t,\varphi )$
and $\Psi^* (t,\varphi +2\pi)=-\Psi^* (t,\varphi )$. Such a field
can be expanded as
\be \Psi (t,\varphi )\ =\ \sum_r \xi_r(t)e^{ir\varphi}\ ,\
\Psi^*(t,\varphi )\ =\ \sum_r \xi^*_r(t) e^{-ir\varphi}\
.\label{IV.12a}\ee
Here $r$ is half-integer, $\xi_r= \xi_r(t)$ and $\xi^*_r=
\xi^*_r(t)$ are anticommuting variables. Below, we consider only
the field action on a circle with radius $\rho$ for left-movers:
\[ S[\Psi^*,\Psi]\ =\ {1\over 2\pi\rho}\int_{{\bf R}\times S^1} dt
d\varphi\, [{i\over 2}\rho(\Psi^*{\dot \Psi}-{\dot\Psi}^*\Psi
)+i\Psi^*\partial_\varphi\Phi ]\]
\be \pm\ {i\over 4\pi}\int_{S^1}d\varphi\,
[\Psi_f^*\Psi_f+\Psi_i^*\Psi_i] \ .\label{IV.13}\ee (the action
for right-movers is obtained by replacement $+i\partial_\varphi\to
-i\partial_\varphi$). This system contains first class constraints
which lead to the Dirac brackets:
\[ \{ \Psi (t,\varphi ),\Psi (t,\varphi' )\}_D\ =\ \{
\Psi^*(t,\varphi ), \Psi^* (t,\varphi' )\}_D\ =\ 0\ ,\]
\be \{ \Psi (t,\varphi ),\Psi^*(t, \varphi' )\}_D\ =\ 2\pi\delta
(\varphi -\varphi' )\ .\label{IV.14}\ee
The +/- signs in (\ref{IV.13}) refer to the cases with the
following fixed values of final and initial fields:
\be +\ {\rm sign:}\ \ \Psi_i(\varphi )\ =\ \sum_r \xi_{ir}
e^{ir\varphi}\ ,\ \Psi^*_f (\varphi )\ =\ \sum_r \xi^*_{fr}
e^{-ir\varphi}\ ,\label{IV.15}\ee
\be -\ {\rm sign:}\ \ \Psi_f (\varphi )\ =\ \sum_r
\xi_{fr}e^{ir\varphi}\ ,\ \Psi^*_i(\varphi )\ =\ \sum_r
\xi^*_{ir}e^{-ir\varphi}\ .\label{IV.16}\ee
Inserting the expansions (\ref{IV.12a}) into (\ref{IV.13}), we
obtain the action
$$ S[\Psi^*,\Psi]\ =\ \int dt\sum_r\, [{i\over
2}\xi^*_r(t){\dot\xi}_r(t)-{i\over 2}{\dot\xi}^*_r(t) \xi_r(t)$$
\be -{r\over\rho}\xi^*_r(t)\xi_r(t)]\ \pm\ {i\over 2}\sum_r\,
[\xi^*_{fr}\xi_{fr} +\xi^*_{ir}\xi_{ir}]\ ,\label{IV.17}\ee
describing the system of independent fermionic oscillators with
frequencies $\omega =r/\rho$.

For (\ref{IV.15}) the solution of equations of motion
\[  \xi_r(t)\ =\ e^{-ir(t-t_i)/\rho} \xi_{ir}
\ ,\ \xi^*_r\ =\ e^{-ir(t_f-t)/\rho}\xi^*_{fr}\ ,\]
gives the Hamilton principal function
\be S_c(\Psi^*_f,\Psi_i)\ =\ i\sum_r\, e^{-ir(t_f-t_i)
/\rho}\xi^*_{fr} \xi_{ir}\ .\label{IV.18}\ee
Similarly, for (\ref{IV.16}) the solution
\[ \xi_r(t)\ =\ e^{ir(t_f-t)/\rho} \xi_{fr}\ ,\
\xi^*_r\ =\ e^{ir(t-t_i)/\rho} \xi^*_{ir}\ ,\]
induces the Hamilton principal function
\be S_c(\Psi^*_i,\Psi_f)\ =\ i\sum_r\,
e^{ir(t_f-t_i)/\rho}\xi_{fr}\xi^*_{ir}\ .\label{IV.19}\ee

In the discrete-time approach, the field modes are described by
fermionic variables $\xi^r_n \equiv\xi^r(n\tau)$ for $n$-odd, and
$\xi^{r*}_{n'}\equiv\xi^{r*}(n'\tau)$ for $n'$-even. Motivated by
(\ref{IV.18}) and (\ref{IV.19}) we take the time-slice system
function in the form
\[ S_\tau\ =\ \sum_r\sum_n (\xi_{n+1}^{r*}\xi_n^r
e^{-ir\lambda}-\xi_{n-1}^{r*}\xi_n^r e^{ir\lambda}) \]
\be =\ \sum_r\sum_{n'}(\xi_{n'}^{r*}\xi_{n'-1}^r
e^{-ir\lambda}-\xi_{n'}^{r*} \xi_{n'+1}^r e^{ir\lambda})\
.\label{IV.20}\ee

After quantization the discrete time analogs of Dirac brackets
(\ref{IV.14}) are replaced by the anticommutation relations
\[ [\Psi_n(\varphi ),\Psi_n(\varphi' )]\ =\ [\Psi^*_{n'}(\varphi ),
\Psi^*_{n'} (\varphi' )]\ =\ 0\ ,\]
\be [\Psi_n(\varphi ),{1\over 2} (\Psi^*_{n+1}(\varphi' )+
\Psi^*_{n-1}(\varphi' ))]\ =\ 2\pi\delta (\varphi -\varphi' )\
.\label{IV.21}\ee
Here, we have taken into account the fact that $\Psi_n(\varphi )$
is affiliated with $n$-odd, and $\Psi^*_{n'}(\varphi )$ with
$n'$-even, and again we have replaced $\Psi^*_{n'}(\varphi )$ by
the nearest neighbor average.

Expanding the fields $\Psi_n(\varphi )$ and $\Psi^*_{n'}(\varphi
)$ into solutions of equations of motion
\[ \xi_n^r\ =\ {1\over\cos r\lambda}e^{-inr\lambda}b_r\ ,\
\xi^{r*}_{n'}\ =\ {1\over\cos r\lambda}e^{in'r\lambda}b^*_r\ ,\ \
r>0\ ,\]
\be \xi_n^r\ =\ {1\over\cos r\lambda}e^{-inr\lambda} {\tilde
b}^*_{-r}\ ,\ \xi^{r*}_{n'}\ =\ {1\over\cos r\lambda}
e^{in'r\lambda}{\tilde b}_{-r}\ ,\ \ r<0\ ,\label{IV.22}\ee
it can be easily seen that the anticommutation relations
(\ref{IV.21}) are satisfied provided $b_r$, $b^*_r$, ${\tilde
b}_r$ and ${\tilde b}^*_r$ satisfy the anticommutation relations
\be [b_r ,{\tilde b}_{r'} ]\ =\ 0\ ,\ [b_r ,b_{r'} ]\ =\ [{\tilde
b}_r ,{\tilde b}_{r'} ]\ =\ \cos (\lambda\tau)\, \delta_{r+r',0}\
.\label{IV.23}\ee
Here we put $b_{-r}=b^*_r$ and ${\tilde b}_{-r}={\tilde b}^*_r$.
For a given $\lambda$ the admissible values of $r$ are specified
by the inequality $-{\pi\over 2\lambda}<r<+{\pi\over 2\lambda}$.

In the Euclidean discrete-time case the anticommutation relations
(\ref{IV.23}) are replaced by
\be [b_r ,{\tilde b}_{r'} ]\ =\ 0\ ,\ [b_r ,b_{r'} ]\ =\ [{\tilde
b}_r ,{\tilde b}_{r'} ]\ =\ [r]_+\, \delta_{r+r',0}\
,\label{IV.24}\ee
where $[r]_+=\cosh r\lambda$ (there is no restriction on
admissible values of $r$). Thus, we obtain two independent sets of
fermionic oscillators both satisfying {\it exactly} the
anticommutation relations required for the deformation in question
\cite{BC}. Any of them can be used for a fermionic realization of
deformed Virasoro algebra and its supersymmetric extension.

The deformed Virasoro algebra generators are usually expressed in
terms of an auxiliary Euclidean real field $\psi(z)$ depending,
for right-movers, on the variable $z=e^{n\lambda +i\varphi}$. This
field can be determined as follows. We restrict ourselves to modes
with $r>0$. The Euclidean field solutions then possess the
expansions (into functions (\ref{IV.22}) with $\lambda$ replaced
$-i\lambda$):
\[ \Psi_n (\varphi )\ =\ \sum_{r>0}{1\over \cos
r\lambda}e^{-inr\lambda+ir\varphi}b_r\ ,\ n\ -\ {\rm odd}\ ,\]
\be \Psi^*_{n'}(\varphi )\ =\ \sum_{r>0}{1\over\cos
r\lambda}e^{in'r\lambda-ir\varphi}b^*_r\ ,\ n'\ -\ {\rm even}\
.\label{IV.25}\ee
In order to construct $\psi(z)$ we take the proper discrete-time
nearest neighbor combinations:
\[ {1\over 2}(\Psi_n(\varphi +\lambda )+\Psi_n (\varphi -\lambda))\ +\
{1\over 2}(\Psi^*_{n+1}(\varphi )+\Psi^*_{n-1} (\varphi )) \]
\be \ =\ \sum_{r>0}e^{(nr\lambda -r\varphi)} b_r\ +\ \sum_{r>0}
e^{i(nr\lambda -r\varphi)}b^*_r\ ,\ n-{\rm odd}\ ,\label{IV.26}\ee
\[ {1\over 2}(\Psi^*_{n'}(\varphi +\lambda )+\Psi^*_{n'}(\varphi
-\lambda ))\ +\ {1\over 2}(\Psi_{n'+1} (\varphi )+\Psi_{n'-1}
(\varphi ))\] \be \ =\ \sum_{r>0} e^{i(n'r\lambda -r\varphi)}
b^*_r\ +\ \sum_{r>0} e^{-i(n'r\lambda -r\varphi)/} b_r\ ,\ n'-{\rm
even}\ .\label{IV.27}\ee
Equations (\ref{IV.26}) and (\ref{IV.27}) define the real
Euclidean field
\be \psi(z)\ =\ \sum_r\, b_r z^{-r}\ ,\ z=e^{n\lambda +i\varphi}\
,\ n-{\rm integer}\ .\label{IV.27a}\ee
Repeating the same procedure for modes with $r<0$, we obtain an
other auxiliary field ${\tilde\psi}({\bar z})$ in terms of
${\tilde b}_r$.

The deformed Virasoro algebra generators in the fer\-mionic
realization are given in terms of $\psi (z)$ as follows:
\[ H^k_n\ =\ \oint{dz\over 2\pi i}\, z^n{1\over 2\lambda}:\psi
(e^{k\lambda/2}z)\psi (e^{-k\lambda/2}z): \]
\be =\ {1\over 2} \sum_r[k({n\over 2}-r)]_- :b_r b_{n-r}:\
.\label{IV.28}\ee
They satisfy, up to central term, the same commutation relations
as $B^k_n$ given in terms of $\Pi (z)$ in (\ref{III.8}). Both
$B^k_n$ and $H^k_n$ span the even part of the deformed super
Virasoro algebra, whereas the odd generators are given in terms of
$\psi (z)$ and $\Pi (z)$:
\[ G^k_r \ =\ \oint{dz\over 2\pi i}\, z^r\psi (e^{k\lambda/2}z) \Pi
(e^{-k\lambda/2}z) \]
\be =\ \sum_j e^{-k({r\over 2}-j)\lambda} a_j b_{r-j}\
.\label{IV.29}\ee
The graded commutator relations generated by $\{B^k_n,H^k_n,G^k_n
\}$ can be found, e.g. in \cite{KS} (our definition of $[k]_-$
differs from that given there by factor $\kappa
=\sqrt{(1/\lambda)\sinh\lambda}$, $[k]_+$ is the same;
consequently, the bosonic oscillators $a_k$ used above are
multiplied by constant $\kappa$ with respect to those used in
\cite{KS}, the fermionic oscillators $b_r$ are unchanged).

\section{Concluding remarks}

The deformed Virasoro algebra \cite{CP1} and its supersymmetric
extension \cite{BC} were suggested earlier on purely formal
mathematical grounds. The field theoretical origin of the deformed
(super) Virasoro algebras, formulated above in the framework of
the Euclidean discrete-time QFT, can serve for a better (physical)
motivation and understanding of its role in all related
constructions ($q$-strings, $q$-vertex operators and
Zamolodchikov-Faddeev algebras).

In this context it would be of great interest to extend our model
to the deformed Kac-Moody algebras, see e.g. \cite{Re} and refs
therein. It is plausible that this can be achieved along the same
lines as in the Virasoro algebra case: (i) The Kac-Moody algebras
can be defined as free field current algebras on a circle; (ii)
Their deformations are realized in terms of various sets of
deformed bosonic and/or fermionic oscillators. As other possible
application could serve discrete-time integrable models, see
\cite{BBR}.

{\it Acknowledgments.} The financial support of the Aca\-demy of
Finland under the Projects No. 54023 is greatly acknowledged.
P.P.'s work was partially supported also by VEGA project
1/7069/20.

\end{document}